\documentclass{czjp}
\usepackage{epsf}  
\usepackage{epsfig}  
\usepackage{amssymb}

\newcommand{\cp}{\cos \psi^*}

\newcommand{\sinw}{\sin^2\theta_W}
\newcommand{\ptm}{P_{\tau}}

\def\tenn{\tau^- \rightarrow e^- \bar{\nu_e}\nu_\tau}
\def\tmnn{\tau^- \rightarrow \mu^- \bar{\nu_\mu}\nu_\tau}
\def\tll{\tau^- \rightarrow l^- \bar{\nu_l}\nu_\tau}
\def\tpn{\tau^- \rightarrow \pi^- \nu_\tau}

\def\trn{\tau^- \rightarrow \rho^- \nu_\tau}
\def\trh{$\tau^- \rightarrow \rho^- \nu_\tau\ $}

\def\taone{\tau^- \rightarrow a_1^- \nu_\tau}

\newcommand{\ptau}{{\ifmmode {P_\tau} \else ${P_\tau}$\fi}}
\newcommand{\rh}{{\ifmmode {\rho\ } \else $\rho\ $\fi}}
\def\mdp{m_\pi^2}
\def\mdt{m^2_\tau}
\def\pc{$\pi\ $}

\def\tnu{\mbox{$\nu_\tau$ }}

\newcommand{\be}{\begin{equation}}
\newcommand{\ee}{\end{equation}}
\newcommand{\bea}{\begin{array}}
\newcommand{\ena}{\end{array}}
\newcommand{\beqn}{\begin{eqnarray}}
\newcommand{\eeqn}{\end{eqnarray}}

\newcommand{\xxe}{\begin{equation}}
\newcommand{\yye}{\end{equation}}
\newcommand{\xxa}{\begin{array}}
\newcommand{\yya}{\end{array}}
\newcommand{\xxeqa}{\begin{eqnarray}}
\newcommand{\yyeqa}{\end{eqnarray}}
\newcommand{\xxeqan}{\begin{eqnarray*}}
\newcommand{\yyeqan}{\end{eqnarray*}}

\def\m{M_Z}
\newcommand{\bi}{\begin{itemize}}
\newcommand{\ei}{\end{itemize}}
\newcommand{\xxitm}{\begin{itemize}}
\newcommand{\yyitm}{\end{itemize}}

\newcommand{\xxfig}{\begin{figure}}
\newcommand{\yyfig}{\end{figure}}
\newcommand{\xxtab}{\begin{table}}
\newcommand{\yytab}{\end{table}}
\newcommand{\xxtabu}{\begin{tabular}}
\newcommand{\yytabu}{\end{tabular}}
\newcommand{\xxc}{\begin{center}}
\newcommand{\yyc}{\end{center}}
\newcommand{\xxmin}{\begin{minipage}}
\newcommand{\yymin}{\end{minipage}}
\newcommand{\xxpic}{\begin{picture}}
\newcommand{\yypic}{\end{picture}}
\newcommand{\xxtop}{\begin{flushright}}
\newcommand{\yytop}{\end{flushright}}
\newcommand{\xxltop}{\begin{flushleft}}
\newcommand{\yyltop}{\end{flushleft}}
\newcommand{\xxquo}{\begin{quote}}
\newcommand{\yyquo}{\end{quote}}

\def\flow{\mbox{$\rightarrow$}}

\def\mt{m_\tau}

\def\t2t{\mbox{$\tau\tau$}}
\def\g2g{\mbox{$\gamma\gamma$}}
\def\m2m{\mbox{$\mu\mu$}}

\def\p0i{\mbox{$\pi^\circ$ }}

\def\tok2pz{\mbox{$\tm \flow \tnu$ K$^-\pi^\circ \pi^\circ$}}

\newcommand{\fp}{{f^+}}
\newcommand{\fm}{{f^-}}
\newcommand{\fpm}{{f^\pm}}

\newcommand{\tpm}{{\tau^\pm}}

\newcommand{\tp}{{\tau^+}}
\newcommand{\tm}{{\tau^-}}
\begin{document}

\begin{frontmatter}
\title{SPIN PHYSICS AT $e^+e^-$ COLLIDERS}
\author{Walter Bonivento}
\address{Laboratoire de l'Acc\'el\'erateur Lin\'eaire\\
 IN2P3-CNRS et Universit\'e de Paris-Sud \\
 BP34, F-91898 Orsay, France}
%
%
%
%
\begin{abstract}
A large number of  measurements with polarized beams and/or spin analysis of 
final state particles  has been performed  at the $e^+e^-$ 
colliders LEP and SLC,  providing important information on the dynamics of  high energy interactions.
In this paper three subjects, for which  the role of spin studies was particularly relevant, will be covered: the measurements of the electroweak couplings, the study of fragmentation dynamics and the search for  physics beyond the Standard Model.
\end{abstract}

\end{frontmatter}

\section{Electroweak physics}

The measurement of the electroweak couplings of the Z to the fermions has been one of the main goals of the high energy colliders LEP (during the first six years of running) and SLC.  \\
Spin effects provide  sensitive measurements of the electroweak couplings and allow to perform accurate tests of the Standard Model. \\
 From the experimental point of view,   high energy 
$e^+e^-$ colliders allow  
to select the different 
final states with high efficiency and low background. \\

\subsection{ The $e^+e^-$ reaction at LEP1} 

 The differential cross-section for  
 fermion-antifermion production, \mbox{$e^+e^-\to f\bar{f}$},
{ in the Standard
Model} (or more generally in a model with chirality-conserving interactions)
can be written, { at tree level, around the $Z$ resonance} and for unpolarized $e^\pm$ beams, as~\cite{Alem92,Berna91}:
\be
\frac{d\sigma}{d\Omega}=\frac{1}{4\pi^2 q^2}|P(q^2)|^2(A+B_{1\mu}s^\mu_{\fp}+B_{2\mu}s^\mu_{\fm}+
C_{\mu\nu}s^\mu_\fp s^\nu_\fm)
\label{mele}
\ee
where  $P(q^2)$ is the Z propagator, $s_\fp^\mu$ and $s_\fm^\mu$ the covariant spin vectors of the outgoing fermions. 
{ Only the $Z$ exchange} contribution is included, being dominant over the 
other ones. \\
The spin-independent term  is
\be
{ A}=C_0(1+cos^2\theta)+2C_1cos\theta
\label{linear}
\ee
 with 
$C_0=(|v_e|^2+|a_e|^2)(|v_f|^2+|a_f|^2)$ and $ C_1=4 Re(v_ea_e^*)Re(v_f a_f^*)$
where  $\theta$ is the outgoing fermion $f$ polar angle with respect to the $e^-$ direction, 
  $v_f$ and $a_f$, $f=e,\mu,\tau$, are the electroweak vector and axial 
coupling constants (for simplicity radiative corrections will be neglected in 
the  following).  The second term of Eq.(\ref{linear}) gives rise to a
forward-backward asymmetry of the cross-section. \\ 
Using  the coordinate system in the laboratory frame where the 
$z$ axis is along the outgoing fermion $f$ direction and the $y$ axis is normal to the reaction plane, 
at leading order in $m_f$:
\be
{ B_{1\mu}s^\mu_\fp+B_{2\mu}s^\mu_\fm }= 
  -D_0(s^{*z}_{\fp}+s^{*z}_{\fm})(1+cos^2\theta)-
D_1(s^{*z}_{\fp}+s^{*z}_{\fm})2cos\theta
\label{line}
\ee
where $\mathbf{s}_\fpm^*$ designates the polarization vector in the $f^\pm$ rest frame. \\
The two terms of Eq.(\ref{line}) give rise to the  { mean longitudinal final-state polarization asymmetry} (a parity(P)-odd observable)
\be <P_f>=-\frac{D_0}{C_0}=-2\frac{Re(v_f a_f^*)}{|v_f|^2 + |a_f|^2}\equiv -{\cal A}_f 
\label{pf}
\ee and
to the  forward-backward polarization asymmetry (P-odd)
\be A_{pol}^{FB}=-\frac{3}{4}\frac{D_1}{C_0}=
-\frac{3}{4}\cdot\frac{Re(v_e a_e^*)}{|v_e|^2+ |a_e|^2}=-\frac{3}{4}{\cal A}_e
\label{pz}
\ee
The latter observable is also related to the  { $Z$ polarization} as 
$P_Z=-{\cal A}_e$.
Other terms related to  transverse
spin components are suppressed by $m_f/M_Z$. 
Using the notation of Eqs.(\ref{pf},\ref{pz}), 
the spin-independent term of Eq.(\ref{mele}) can  be rewritten as 
\be 
A \propto (1+cos^2\theta)-2 {\cal A}_f  P_Z cos\theta
\label{afb}
\ee
The quadratic term in the spin of Eq.(\ref{mele}) gives rise to the spin
correlations:
\be
{ C_{\mu\nu}s^\mu_\fp s^\nu_\fm}=C_0h_0+C_1h_1+C_2h_2+D_2h_3
\ee
with
\beqn
h_0 & = & s^{*z}_{\fp}s^{*z}_{\fm}(1+cos^2\theta) \\
h_1 & = & s^{*z}_{\fp}s^{*z}_{\fm}2cos\theta \\
h_2 & = & (s^{*y}_{\fp}s^{*y}_{\fm}-s^{*x}_{\fp}s^{*x}_{\fm})sin^2\theta \\ \label{tran}
h_3 & = & (s^{*y}_{\fp}s^{*x}_{\fm}+s^{*x}_{\fp}s^{*y}_{\fm})sin^2\theta
\eeqn
The terms $h_0$ and $h_1$ represent the { longitudinal spin correlation},
 $h_2$   the { transverse-transverse spin correlation}, whose coefficient is 
$ C_{TT}\equiv C_2/C_0=(|a_f|^2-|v_f|^2)/(|a_f|^2+|v_f|^2)$
and $h_3$ the { transverse-normal spin correlation, with coefficient 
$C_{TN}$} (P-odd and 
T-odd).
In the Standard Model, at tree level, the values of the  Z couplings to fermions  depend on one parameter, the Weinberg angle $\theta_W$. 
For $\sinw=0.232$,  $ {\cal A}_l=0.15$,
${\cal A}_u=0.67$ and ${\cal A}_d=0.935$.  The { longitudinal spin correlation} is one by helicity conservation and, for leptonic final states, 
$C_{TT}=0.99$ and $C_{TN}=-0.01$~\cite{trasAL97}.

\subsection{ The $\tau$ decay as polarization analyser}
 The $\tau$ spin vectors are not directly observable. The $\tau$ weak decay products can be used as $\tau$ { spin analyser}.  
In the  $\tau^\pm$ rest frame  
\be 
d\Gamma_X^\pm ({\mathbf{s}}^*_\tpm)
\propto\left(1\pm\alpha_X m_\tau 
\frac{(\mathbf{q}_\tpm^*\cdot \mathbf{s}_\tpm^*)}{(\mathbf{q}_i^*\cdot 
\mathbf{p}_\tpm)}\right)d\Omega_\tpm^*
\label{decay}
\ee
where $\mathbf{q}_\tpm^*$ is the  momentum of the $\tpm$ decay product, $\mathbf{p}_\tpm$
  is the  momentum of the $\tpm$ in the laboratory frame, 
 $\Gamma_X^\pm$ is the $\tpm$ partial decay width in the decay channel $X$.
 $\alpha_X$ is a constant whose value depends on the decay channel $X$ and is called  {\em analysing power}.
By CP invariance $P_{\tau^+}=-P_{\tau^-}$; from Eq.(\ref{decay}) it is seen  that particle and anti-particles have the same decay spectra. By definition it will be taken $P_\tau\equiv P_{\tau^-}$.
\subsection{ Measurement of the $\tau$ longitudinal polarization} 
The $\tau$ polarization was measured from the study of single 
$\tau$ decay products spectra. A method based on the acollinearity spectrum
of the $\tau$ decay products was also used, though it was found to be less
 sensitive to the polarization. It was also
 shown~\cite{Davier93,tpolAL98} that, 
 when both $\tau$'s decay semileptonically, an optimal sensitivity can be obtained by reconstructing the $\tau$ direction. \\
For what concerns their decay spectra, semileptonic  and leptonic decays have to be treated in a different way.
\subsubsection{ Spin-0 hadronic system ($\tau\to \pi(K) \nu_\tau$)} 
There are two  amplitudes for the decay, corresponding to the two possible $\tau$ polarization states: $A^+\propto\cos{\theta^*/2}$ and $A^-\propto\sin{\theta^*/2}$, where $\theta^*$ is the angle in the $\tau$ rest frame  of the hadron momentum with respect to  the polarization axis of the $\tau$ (defined as its direction of motion in the laboratory frame). Then:
\be
W(cos\theta^*)={dN\over dcos\theta^*}
\propto \frac{1+\ptau}{2}|A^+|^2 +\frac{1-\ptau}{2}|A^-|^2 =[1+\ptau cos\theta^*]
\ee
The analysing power, defined as in Eq.(\ref{decay}), for this decay is
 $\alpha_\pi=- 1$.
In the two body decay of the $\tau$ the angle $\theta^*$ is fully correlated to the hadron energy  in the lab frame: $cos\theta^*\simeq 2x_\pi -1$ with  
$x_\pi=(E_{\pi}/E_{beam})_{LAB}$. Therefore more  energetic hadrons correspond to
 $ P_{\tau}=+1$.

\subsubsection{ Spin-1 hadronic system ($\tau\to\rho(K^*)\nu_\tau$ , $\tau\to a_1\nu_\tau$)}
In this case there are  four decay amplitudes, for the two allowed 
 hadronic system helicities $\lambda_X=0,-1$, $X=\rho,a_1$ and 
$ W(cos\theta^*)\propto 1+\ptau\alpha_X cos\theta^* $,
with $\alpha_X=(\mt^2-2m_X^2)/(\mt^2+2m_X^2)$.
With respect to the pion decay, the analysing power is reduced:
 $\alpha_\rho =- 0.46$ for \trh and $\alpha_{a_1}=-0.12$  for $\taone$. \\
To { regain  sensitivity} it is possible to measure the hadron spin composition
from the spectra of its  decay products.
For the decay \trh, in the $\rh$ frame, 
${dN(\rh_T\to\pi\pi)/dcos\psi^*}\propto sin^2\psi^*$ and $ {dN(\rh_L\to\pi\pi)/ dcos\psi^*}\propto cos^2\psi^*$, 
where $\rho_T$ and $\rho_L$ indicate a transversely and longitudinally 
polarized $\rho$, respectively,  and $\psi^*$  is the 
 angle, in the $\rho$ frame,   of the charged \pc with respect to  the polarization axis of the $\rho$.
Like $\theta^*$, $\psi^*$ can also be reconstructed from measured particle energies
in the lab system as $\cp=
(2x_\rho -1)\sqrt{1- 4\mdp/\mdt}$ with 
 $x_\rho=(E_\pi/E_\rho)_{LAB}$.
Most of the information on the $\tau$ polarization is therefore contained in
 the two-dimensional distribution $W(\theta^*,\psi^*)$. \\ 
An optimal one-dimensional variable, $\omega$,
with maximal sensitivity to the $\tau$ polarization was derived in~\cite{Davier93}. Indeed, given $ W(\vec{x})\propto f(\vec{x})+P_\tau g(\vec{x})$, where $\vec{x}$ is 
the vector whose  components are the  variables  which the distribution depends on, the optimal variable is $\omega=g(\vec{x})/f(\vec{x})$. \\
For the decay $\taone$ ,$a_1\to 3\pi$ similar arguments apply, though 
there is a residual $a_1$ decay  model dependence which cannot be factorized out and remains as a systematic uncertainty on $P_\tau$~\cite{Privi93}.

\subsubsection{ Leptonic decays} 

{ In the $\tau$ frame} $dN/(dx_l^* dcos\theta^*)\propto  x_l^{2*}[(3-2x^*_l)+
\ptau(1-2x^*_l)cos\theta^*]$,
where $x^*_l=2 E^*_l/\mt$ and $E^*_l$ is the charged lepton energy.
  For leptonic decays $ \alpha_l= 1/3$, of opposite sign with respect to  hadronic decays. In the laboratory system,
 contrary to the hadronic case, $P_{\tau}=-1$ gives more energetic leptons. \\
The sensitivity to the longitudinal polarization for a given $\tau$ decay 
channel~$X$ is defined as\linebreak $S_X\equiv 1/(\Delta \ptm^X \sqrt{N})$,
where N is the number of selected events used to obtain 
an error $ \Delta \ptm^X$ on the polarization. Table~\ref{table} lists the different sensitivities for the various  $\tau$ decay channels, in the case of an ideal detector. \\
\begin{table}[h]
\begin{center}
\begin{tabular}{|l|c|c|c|} \hline
 decay channel $X$ & $S_X$ & $B_X$ & $w_X$ \\ \hline  \hline
 $\tpn$  & 0.58 & 0.12 & 1 \\
 \trh & 0.49 ($0.58^*$) & 0.26 &  1.6  \\
 $\taone$  & 0.45 ($0.58^*$) & 0.10 & 0.5  \\
 $\tenn$ &  0.22 &  0.18 & 0.2  \\
 $\tmnn$ &  0.22  & 0.17 & 0.2  \\ \hline
\end{tabular}
\end{center}
\caption{Sensitivity $S_X$ for an ideal detector to the $\tau$ polarization 
for the decay channel $X$~\protect~\cite{Davier93}, branching ratio $B_X$ and relative weight 
$w_X=S^2_X B_X$ in the average of the results of the different decay 
channels~\protect\cite{PDG98}, normalised to the pion value ($*=$ using the
reconstruction of the $\tau$ direction~\protect\cite{tpolAL98}).}
\label{table}
\end{table}

\newpage
The measurement of  $P_\tau$ as a function of $\theta$ allows to 
 extract at the same time ${\cal A}_\tau$ and $P_Z$:
\be P_\tau(cos\theta,P_e)=
-\frac{
(1+cos^2\theta){\cal A}_\tau-2cos\theta P_Z}{
(1+cos^2\theta)-2cos\theta{\cal A}_\tau P_Z}
\label{ptauang}
\ee

\subsection{ Measurement of the $\tau^+\tau^-$ transverse spin correlations} 

\begin{figure}[htbp]
\begin{center}
\mbox{\epsfig{file=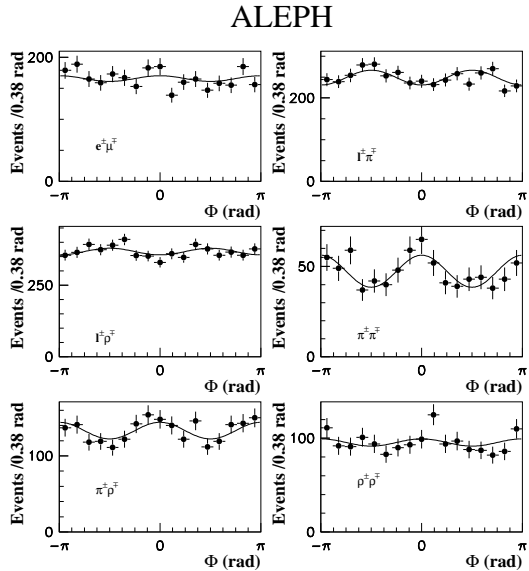,height=80mm}}
\end{center}
\caption{Transverse spin correlations: aplanarity angle, as defined in the text,  distributions. The curve shows the Standard Model prediction normalised to the total number of events~\protect\cite{trasAL97}.}
\label{tra}                                               
\end{figure}

Introducing the $\tau$ decay  in Eq.(\ref{mele})
one obtains:
\be
\frac{d\sigma(X_+,X_-)}{d{ \epsilon} dcos\theta_- d{\phi}}  =  
 \frac{|P(q^2)|^2}{4q^2}
(F_0({ \epsilon})(1+cos^2\theta_-) 
 +  F_1({ \epsilon})2cos\theta_-+
F_2 ({ \epsilon},{ \phi})sin^2\theta_- )
\ee
where
\beqn
F_n({ \epsilon}) & = & C_n [Q_1({\epsilon})+\alpha_{X_+}\cdot\alpha_{X_-}
 Q_2({ \epsilon})] +D_n (\alpha_{X_+}-\alpha_{X_-}) Q_3({\epsilon}) \nonumber \\ 
F_2({\epsilon},{\phi}) & = & \alpha_{X_+}\cdot\alpha_{X_-} [C_2 cos2{ \phi}+D_2 sin2{\phi}]
 Q_4({\epsilon})
\label{trasv}
\eeqn
 with $n=0,1$, $\theta_-$  the 
polar angle of $\tau^-$ decay product, $X_+$ and $X_-$ the $\tau^+$ and $\tau^-$ decay channel, respectively, 
 ${ \epsilon}$ the acollinearity angle between the two $\tau$ decay products and  ${ \phi}$ the aplanarity angle, defined as the azimuthal angle of the 
$e^-$ beam axis in a reference frame where the z axis is along the momentum of 
$\tau^-$ decay product and the momentum of 
$\tau^+$ decay product is contained in the $x-z$ plane. The complete expression of the $Q_i(\epsilon)$, $i=1,4$, functions can be found in~\cite{Berna91}. \\
The transverse spin correlations are given by the  terms
$C_2,D_2$ in Eq.(\ref{trasv}). They are  therefore expected to
be observed   as  
 ${\phi}$ modulation of the  cross-section, with an amplitude 
 depending on the product $\alpha_{X+}\alpha_{X-}$. Therefore 
  opposite signs are expected for  hadron-hadron and  hadron-lepton events, respectively.
This was indeed observed~\cite{trasDE97,trasAL97}, as shown in figure~\ref{tra}. One experiment~\cite{trasAL97}  obtained
$C_{TT}=1.06\pm 0.14$ and $C_{TN}=0.08\pm 0.14$, in agreement with the Standard Model predictions.

\subsection{ The $\tau^+\tau^-$ longitudinal spin correlations}
  
Assuming helicity conservation in high energy $e^+e^-$ interactions, the measurement of the longitudinal spin correlations
 can be interpreted as a measurement of the  $\nu_{\tau}$ helicity ($h_{\nu_{\tau}}$). Experiments without beam polarization  measured the absolute value
 of  $h_{\nu_{\tau}}$ using this correlation~\cite{nhellL396,nhellAL96} and both the absolute value and the sign from 
$\taone$ decay spectra, though with modest precision~\cite{aihelOP97}. 

\begin{figure}[h]
\begin{center}
\mbox{\epsfig{file=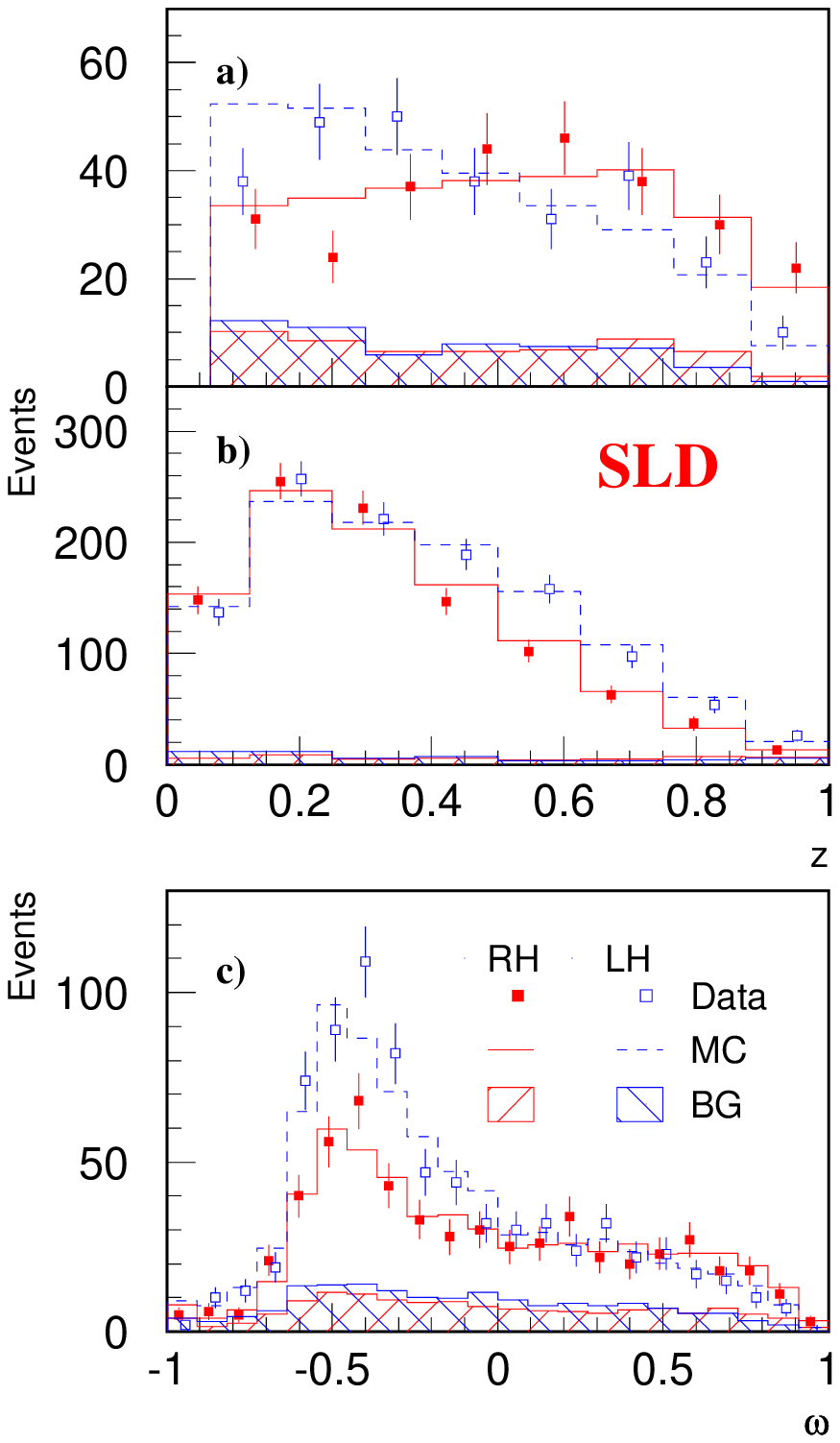,
height=130mm,bbllx=-100,bblly=-100,bburx=467,bbury=467}}
\vspace{-3cm}
\end{center}
 \caption{$\tau$ decay spectra for  a) $\tpn$ and  b) $\tll$, where $z$ is the ratio of the 
charged $\tau$ decay product energy to beam energy in the laboratory frame and for  \break c) $\trn$, with  $\omega$  defined as in~\cite{Davier93}. 
The dots and lines classified as L.H.(left-handed) correspond to the selection  $(cos\theta>0,P_e<0)$ and $(cos\theta<0,P_e>0)$: the  $\tau^-$ is expected to be predominantly left-handed and therefore the $\pi$ and $\rho$ spectra are expected to be softer and the lepton spectrum is expected to be harder, as  observed in the data. From~\protect\cite{nhellSLD97}.}
\label{slc}                                                      
\end{figure}

 At SLC a 
  highly longitudinally polarized electron beam, $|P_{e^-}|>  75\%$, 
collides with an unpolarized positron beam, 
giving rise to { highly longitudinally polarized $\tau$'s}. Indeed, 
Eq.(\ref{ptauang}) is still valid with 
\be
P_Z=-\frac{{\cal A}_e-P_e}{1-{\cal A}_e P_e} 
\label{pzb}
\ee
Therefore $ P_Z(P_e^->0)=0.70$ and $P_Z(P_e^-<0)=-0.83$ while $P_Z(P_e^-=0)=-0.16$.
With appropriate combinations of  beam polarization 
and angular  regions in $\theta$, it is possible to measure $\tau$ spectra which are sensitive to both the magnitude and the sign of the neutrino helicity as
 shown in figure~\ref{slc}. In~\cite{nhellSLD97} {$ h_{\nu_{\tau}}=-0.94\pm 0.09$} was obtained.\\ 

\subsection{ Measurements of the Z couplings at SLC}

At SLC, with beam polarization $P_e^-$, the spin-independent term 
of Eq.(\ref{mele}) has the same form of Eq.(\ref{afb}), though with $P_Z$ defined as in Eq.(\ref{pzb}).
A larger forward-backward asymmetry than at LEP is then expected, resulting in a sensitivity higher by more than a factor 4 to the ${\cal A}_f$ measurement. \\
The possibility at SLC to invert the  beam polarization allows
 to define an observable which is not accessible at LEP, 
the left-right asymmetry:
 \be A_{LR}  =  \frac{\sigma_L-\sigma_R}{\sigma_L+\sigma_R}  = 
 \frac{1}{|P_e|}\cdot \frac{N_L-N_R}{N_L+N_R}  =  {\cal A}_e 
\ee
where $N_L$ ( $N_R$) is the number of selected events with negative (positive) polarization of the electron beam. Due to its linear dependence on
${\cal A}_e $ and to the relatively high statistics that can be collected with an inclusive e$^+$e$^- \rightarrow q \bar q$ selection, the left-right asymmetry gives a very precise measurement of ${\cal A}_e $. \\
\subsection{Extraction of the Weinberg angle}
In the context of the Standard Model, from the different measurements of ${\cal A}_f $,  an estimate of the Weinberg angle was obtained. \\
As  is shown  in a recent review of the subject~\cite{Karlen98}, 
the $\tau$ polarization at LEP and the  left-right asymmetry at SLC 
are among the most sensitive measurements of the Weinberg angle, 
to which they contribute with a relative
 error of 0.18\% and 0.13\%, respectively.

\section{  Fragmentation dynamics}

At LEP/SLC energies  final state quarks and gluons give rise to multi-particle jets in the detectors. Very often studies of spin effects are concentrated on the highest momentum, named leading,  particle of these jets, under the hypothesis 
that the quark initiating the jet is carried by this  particle and  the quark 
  quantum numbers are transferred to it.
A prerequisite for  spin studies is 
  that   the sign of the charge  (taking into account the possible quark combinations in a hadron) 
and the flavour  are transferred from 
the  quark to the leading hadron. \\
 In {  $b\bar{b}$ and $c\bar{c}$ events} B and D hadrons are produced with hard fragmentation functions and at a rate of about two per event. It is therefore plausible that they carry the initial quark and its quantum numbers. \\
For { light quark events}  a measurement  was performed at 
SLC~\cite{leadSLD97}. The electroweak theory predicts that the  
 quark ($q$) jet  should mostly follow the $e^-(e^+)$ beam direction for 
$P_e=-1(+1)$. In this way quark-initiated jets can be selected with  73\%  purity.
The asymmetry $ D_h=(N^+-N^-)/(N^++N^-)$, where $N^+=N(q\to h)+N(\bar{q}\to\bar{h})$,
$N^-=N(q\to \bar{h})+N(\bar{q}\to h)$ and $h$ is the leading hadron in the jet, was measured and found to be significantly different from zero for relatively large momentum fractions of the leading proton, $\Lambda$, $K^{\pm}$ and $\bar{K^{*o}}$. Since baryons do not contain any constituent anti-quark, this indicates
 that  leading baryons are likely to carry the initial quark and its quantum numbers.
For mesons the quantitative interpretation of the data in terms of the static quark  model suggested that fast kaons are likely  to carry the primary quark and its quantum numbers and that leading kaons are produced predominantly in $s\bar{s}$ events rather than in $u\bar{u}$ and $d\bar{d}$ events.
No significant polarization transfer was observed for leading charged pions, though the quark model predicted for them the same polarization transfer as for kaons.
\subsection{ Measurement of the $\Lambda$ and $\Lambda_b$ longitudinal polarization } 
\begin{figure}[h]
\centering
\mbox{\epsfysize=60mm\epsffile{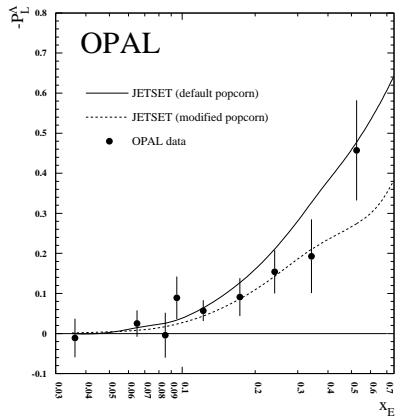}}
\caption{Lambda longitudinal polarization as a function of $x_E=2E_\Lambda/\sqrt{s}$, where $ E_\Lambda$ is the $\Lambda$ energy in the laboratory system,
compared to theoretical predictions.}
\label{fig:lambda}
\end{figure}

At the $Z$ peak the Standard Model predicts for down-type quarks\linebreak
 $<P_L>=-0.91$ for $\sinw=0.232$, including QCD corrections.
{ The main question is whether and to which extent the primary quark polarization is transferred to the leading final state  $\Lambda(\Lambda_b)$ during the hadronization process}.
{ For the  $\Lambda_b$}, HQET predicts~\cite{Mannel91} the decoupling of the heavy quark degrees of freedom  from the light quark ones~\cite{Mannel91}.
 Therefore, full polarization transfer is expected. 
{ For the $\Lambda$}, one has to rely on the constituent model prediction, which is known to be violated in deep inelastic scattering (DIS), though at low 
 $x$ Bjorken. \\
The $\Lambda$ longitudinal polarization was measured~\cite{lamdpOP98,lamdpAL96} using the fact that, in the { parity violating decay}  $\Lambda\to p\pi$, S and P wave final states interfere 
giving rise to the asymmetric angular distribution
  $ W(\theta^*)\propto 1+\alpha_\Lambda P_L cos\theta^*$, where $\theta^*$ is 
the angle of the proton in the $\Lambda$ rest frame, with respect to the polarization axis of the $\Lambda$.
The value  $\alpha_\Lambda=0.642\pm 0.013$ is known 
 from previous experiments~\cite{PDG98}. 
 Since   $\alpha_{\bar{\Lambda}}=-\alpha_\Lambda$, by CP invariance,
and   $\lambda(s)=-\lambda(\bar{s})$, where $\lambda(s)$ is the strange quark helicity, 
 $\Lambda$ and $\bar{\Lambda}$ are expected to give, and it was indeed found in the data, the 
 { same slope}. 
The comparison with  theory was performed taking into account that the final selected sample contained also unpolarized $\Lambda$'s from the decay of other strange baryons ($\Sigma$,$\Xi$,...), $\Lambda$'s  produced during the fragmentation process, $K^o_S$'s  and combinatorial background.
 The two experiments are  consistent with the absence of  depolarizing mechanisms. The result of one of them~\protect\cite{lamdpOP98} is shown in figure~\ref{fig:lambda}. 
\vspace{-0.5cm}

 The  observation of a large polarization transfer will be exploited in 
polarized deep inelastic scattering experiments (see for instance~\cite{compa96}) to test the hypothesis of intrinsic proton strangeness~\cite{Ellis95} and then to shed light on the spin content of the proton. \\

The $\Lambda_b$ longitudinal polarization was extracted 
 from the measurement  of { lepton and neutrino }energy  spectra in the decays
$ \Lambda_b\to \Lambda^+_c l^- \bar{\nu}_l X$, 
$\Lambda^+_c \to \Lambda \pi^+ X$. 
To model the $ \Lambda_b$ decay the HQET prediction was used:
in the $\Lambda_b$ rest frame, the lepton tends to be emitted anti-parallel and the anti-neutrino parallel to the $\Lambda_b$ spin,
giving, for the negatively polarized $\Lambda_b$  in the lab frame, 
 a harder lepton and a softer (anti-)neutrino spectra. 
 The neutrino energy was obtained dividing the event in two hemispheres and  subtracting the hemisphere visible energy to the beam energy. 
A recent average~\cite{VanKooten98} gives 
$ -0.44^{+0.14}_{-0.12}$ and is  consistent with the absence of depolarizing mechanisms, if a significant production
of  $\Sigma_b,\Sigma_b^{(*)}$,
strongly decaying to  $\Lambda_b \pi$, is taken into account.

\subsection{ Spin alignment of  vector mesons  } 

 Information on particle production mechanisms can be obtained from  measurements of  the elements of the helicity density matrix
 $\rho_{\lambda\lambda^{\prime}}$, $\lambda,\lambda^{\prime}=0,\pm 1$ for $J^P=1^-$ vector mesons. The diagonal elements of the matrix represent the probability that the particle has helicity $\lambda$. \\

\newpage
In statistical models  all helicity states of fragmentation quark pairs are 
equally likely to occur. Then, if the spin of primary quark is parallel to that of secondary antiquark, a transversely polarized ($\lambda=\pm 1$)
 vector meson is produced; if it is anti-parallel, a pseudoscalar meson is produced with probability $f$ and a longitudinally polarized ($\lambda=0$) 
vector meson with probability $1-f$. In this model
$\rho_{00}=(1-f)/(2-f)$ and therefore 
$0<\rho_{00}<0.5$. In terms of ratio of pseudoscalar (P) to vector (V) meson production
$\rho_{00}=1/2(1-P/V)$.
A vector meson is said to be spin aligned if $\rho_{00}\not= 1/3$, i.e. if there is a non-uniform population of longitudinal and transverse states, regardless of the values of $\rho_{1,1}$ and $\rho_{-1,-1}$.
For heavy quark resonances, such as $D^*$ and $B^*$, this model, together the HQET prediction of decoupling  of the light quark degrees of freedom, predicts 
 $\rho_{00}=1/3$. 
A QCD-inspired model~\cite{augu80}, describing the fragmentation process 
as the emission of soft gluons from the primary quark predicts  $\rho_{00}=0$ while a model describing vector meson production $V$ through the helicity conserving process 
 $q\to qV$  give~\cite{dono79}
 $\rho_{00}=1$. Finally, in the most popular Lund string model and  QCD cluster model, spin aspects of particle production are essentially ignored. \\
The element $\rho_{00}$ was extracted from the measurement of the resonance production cross-section  
in different  $\theta^*$ intervals,
 where $\theta^*$ is the angle of the meson decay product 
with respect to the meson polarization axis. \\
Two cases were studied: \\
a) { vector meson $\to$ two pseudo-scalars} \\
 The main experimental problems came from 
the presence of  reflections from other resonances (both from imperfect particle identification and from other resonances decaying to the same particles) and 
from resonance shape distortions due to Bose-Einstein effect. \\
An  isotropic distribution is expected for no spin alignment 
($\rho_{00}=\frac{1}{3}$) and proportional to { $sin^2\theta^*$ for helicity $\pm 1$} states and to { $ cos^2\theta^*$ for helicity 0} states. \\
{ Indication for spin alignment at high momentum fraction $x_p$ was obtained at LEP~\cite{kstarDE98,kstar2DE98,kstarOP97,bstarOP97}}.
The combined result of two experiments~\cite{kstarDE98,bstarOP97}, for $\phi\to K K$, $x_p\ge 0.4$ is
$ \rho_{00}=0.49\pm 0.05$ which is $3.2\sigma$ from 1/3.   
The combined result of two experiments~\cite{kstar2DE98,kstarOP97}
  for  $K^{0*}\to K\pi$,  for $x_p\ge 0.7$ is 
$\rho_{00}=0.54\pm 0.06$ which is  $3.46\sigma$ from  1/3.
These results could be consistent with statistical models, though it would require a large suppression of pseudoscalar meson production in the hadronisation.
They are in agreement with the model~\cite{dono79}, 
 which is however 
considered old-fashioned and is no more used in present high energy experiments.   \\
Values consistent with 1/3 were found for $\rho^0  \to  \pi\pi$ and
for $D^{*\pm}  \to D^0\pi\to K\pi\pi $ at all  $x_p$ (though some disagreement among experiments exists) and for 
$\phi\to K K$ and $K^{0*}\to K\pi$ at low $x_p$. \\
b){  vector meson $\to$  pseudo-scalar + vector} \\
The decay  $B^*\to B\gamma$ was studied.
An isotropic distribution is expected 
for no spin alignment and proportional to { $1+cos^2\theta^*$ for helicity $\pm 1$} states and to {  $ sin^2\theta^*$ for helicity 0} states. \\
The combined result of three experiments~\cite{bstarAL96,bstarDE95,bstarOP97}
 is 
$\rho_{00}=0.33\pm 0.04$ which, together with the measured 
$V/(V+P)=0.75\pm 0.10$, is { consistent 
 with a uniform population of  spin states, as predicted by HQET. } \\
Since all these resonances decay by strong or electromagnetic interactions, no separate measurement of  $\rho_{1,1}$ and $\rho_{-1,-1}$ was possible.

\subsection{ Spin composition of $\Lambda$ baryon pairs } 
The  measurement of the { relative polarization of  baryon-antibaryon
 pairs}, selected with   low $Q^2$ in order to remove 
the baryons directly coming  from  the primary  $s\bar{s}$ pair,
allowed to  investigate  the baryon production process. As an example, if the production of these pairs went through single gluon emission, in analogy 
to $J/\Psi$, the S=1 state would be expected to dominate.
For $\Lambda\bar{\Lambda}$ pairs the angular distributions  are~\cite{Alex95}
$dN/dy^*= 1+\alpha_\Lambda^2 y^*$ for S=0  and $dN/dy^*= 
1-\alpha_\Lambda^2/(3 y^*)$ for S=1,
where $y^*=cos \theta^*$ and  $\theta^*$ is the
 angle between the two hyperons'decay protons, 
each measured in its parent decay frame. \\
In the region $Q=\sqrt{M^2_{\Lambda\bar{\Lambda}}-4m^2_\Lambda}<2.5$ GeV, where a broad mass enhancement is observed, it was found~\cite{lalaOP96} that the fraction of the S=1 state contribution is $0.71\pm 0.07$, consistent with a statistical spin distribution, indicating  that $\Lambda\bar{\Lambda}$ pair production goes through   many QCD  processes and baryon's decay.

\section{  Search for New Physics: the $\tau$
 weak magnetic dipole moment}

{ In the S.M. the  weak magnetic dipole moment   of the $\tau$ lepton is
 zero at Born level}. A calculation including 
  higher order corrections gives~\cite{Berna95,Berna94}
\mbox{$a_\tau^w=-(2.10 + i \ 0.61)10^{-6}$}. 
{ A measurement of $a_\tau^w$  significantly different from this prediction would be an evidence for new physics beyond the Standard Model}. \\
In analogy with the e.m. dipole moment, the weak one is introduced using the effective Lagrangian
 \be {\cal L}=\frac{1}{2}\frac{ea_\tau^w }{2m_\tau}\bar{\psi}\sigma^{\mu\nu}\psi Z_{\mu\nu}\ \ \ \
\mathrm{and} \ \ \ \
 \mu_\tau^w=\frac{e}{2m_\tau}2(\frac{1-4sin^2\theta_W}{4sin\theta_W cos\theta_W}+a_\tau^w) \ee 
with $Z_{\mu\nu}=\partial_\mu Z_\nu-\partial_\nu Z_\mu.$
Keeping only up to linear  terms in the spin and in the weak
dipole moment and
neglecting terms proportional to the electron mass,
the tree level  $e^+\, e^- \longrightarrow
\tau^+ \tau^-$ cross-section at the $Z$-peak can be written as the sum of a spin-independent term and a spin-dependent one, which is:
\be
\frac{d \sigma^{S}}{d \Omega_{\tau^-}} \propto
[\  (s^{*x}_{\tp}+s^{*x}_{\tm}) X_++  
 (s^{*y}_{\tp}+s^{*y}_{\tm}) Y_++  
 (s^{*z}_{\tp}+s^{*z}_{\tm}) Z_+ +(s^{*y}_{\tp}-s^{*y}_{\tm}) Y_-]
\label{cross3}
\ee
 The presence of the dipole moment term in the Lagrangian induces therefore, 
in addition to the longitudinal ($Z_+$) polarization term,  
single $\tau$  transverse  ($X_+,Y_+$), P-odd,   and { normal} ($Y_-$), T-odd, polarization terms. Not including terms suppressed by the large value of $\gamma=M_Z/2 m_\tau$, the transverse polarization terms are given by:
\beqn
{ X_+} &=&a_\tau\; \sin\theta_{\tau^-}2\gamma\left[4v_\tau^2 +(v_\tau^2+a_\tau^2)\cos
\theta_{\tau^-}\right] Re(a^w_{\tau})\label{x} \nonumber \\
{ Y_+} &=&\;-2 v_\tau \gamma \beta \sin\theta_{\tau^-}
\;[2 a_\tau^2+(v_\tau^2+a_\tau^2)  \cos \theta_{\tau^-}]\;
 Im(a_{\tau}^w) \label{y+}
\eeqn
Introducing the $\tau$ decay, for the spin-dependent term,  
with both $\tau$'s decaying semi-leptonically, it is  obtained:
\be
\frac{d\sigma^S}{d\cos\theta_{\tau^-}\, d\phi_{h^{\pm}}}\propto 
[\alpha_{h} (\mp X_+\cos\phi_{h^{\pm}}+Y_ \mp\sin\phi_{h^{\pm}})]
\ee
where the reference frame is chosen such as 
$\vec{p}_{\tau^-}$ is along z axis, the $e^-$ beam direction in x-z plane, 
$\theta_{\tau^-}$ is their relative  angle and 
$\phi_{h}$ is the azimuthal angle  of the 
$\tau^-$ decay product momentum in this frame. \\
Using a mixed up-down forward-backward 
angular asymmetry limits were set on the real and imaginary part of 
the weak dipole moment. 
{ To measure $\phi_h$ one needs to  reconstruct the $\tau$ direction.
This is possible with  both $\tau$'s decaying semileptonically and leads to a two-fold ambiguity}. 
It can be solved by measuring the vector $\vec{d}_{min}$ of closest approach 
of the two $\tau$ decay products with the help of a micro-vertex detector. In~\cite{wdipL398} for $\pi-\pi$ events $\epsilon=60-80\%$.
The limits obtained at\linebreak 95\% C.L. were $|Re(a_\tau^w)|<4.5\cdot 10^{-3}$
and  $|Im(a_\tau^w)|<9.9\cdot 10^{-3}$.

\section{Conclusions}

In this paper I presented a number of examples of measurements related to spin
performed during the last ten years at LEP and SLC.
For lack of space, some  important topics have been omitted, such as 
hard QCD studies, i.e. the measurement of the gluon spin, or 
the study of the $W^+W^-Z/\gamma$ vertices.  \\
As a perspective on the future, spin effects  will be 
  very useful   tools for the search for physics beyond the Standard Model at a next higher energy linear collider. Among the most promising 
subjects, there is the study of spin observables in $t\bar{t}$ events,  sensitive to anomalous moments and CP-violating form factors, and the use of 
  polarized beams to reduce backgrounds from standard processes in the search for supersymmetric particles\cite{Haber94}. \\
As an overall conclusion I would say that spin physics is clearly 
a central subject at $e^+e^-$ colliders.

 \section{Acknowledgements}
I would like to thank the organizers of the Workshop for the invitation.
I would also like to thank  A. Kotsinian, L. Serin, and D. Zerwas for their useful comments during the write-up of the paper.



\end{document}